\documentclass[twocolumn,showpacs,preprintnumbers,amsmath,amssymb,prl]{revtex4}

 \def\XXint#1#2#3{{\setbox0=\hbox{$#1{#2#3}{\int}$}
     \vcenter{\hbox{$#2#3$}}\kern-.5\wd0}}

\def\fakebold#1{\relax\ifvmode\leavevmode\fi%
\ifmmode%
\setbox0=\hbox{$#1$}%
\else%
\setbox0=\hbox{#1}%
\fi%
\kern-.02em\copy0 \kern-\wd0%
\kern .04em\copy0 \kern-\wd0%
\kern-.0125em\raise.02em\box0%
}%

\usepackage{graphicx} \usepackage{dcolumn} \usepackage{bm}

\begin{document}

%\preprint{APS/123-QED}

%%%%%%%%%%%%%%%%%%%%%%%%%%%%%%%%%%%%%%%%%%%%%%%%%%%%%%%%%%%%%%%%%%%%%%%%%%%%%%%

\title{Comment on ``Effects of Thickness on the Spin Susceptibility of
  the Two Dimensional Electron Gas''}

\author{Ying Zhang} 
\author{S. Das Sarma} 
\affiliation{Condensed
  Matter Theory Center, Department of Physics, University of Maryland,
  College Park, MD 20742-4111}

\date{\today}

\begin{abstract}
A comment on a recent paper (PRL {\bf 94}, 226405 (2005)) by S. De
Palo {\it et al.}

\end{abstract}


\maketitle

%%%%%%%%%%%%%%%%%%%%%%%%%%%%%%%%%%%%%%%%%%%%%%%%%%%%%%%%%%%%%%%%%%%%%%%%%%%%%%%

In a recent Letter De Palo {\it et al.} ~\cite{depalo} claim that
``Using available quantum Monte Carlo predictions for a strictly 2D
electron gas, we estimate the spin susceptibility of electrons in
actual devices taking into account the effect of the finite transverse
thickness and finding very good agreement with experiments.'' In this
Comment we point out that this claimed ``very good agreement'' is
misleading and accidental because a crucial parameter -- the
background depletion charge density -- determining the quasi-2D
thickness effect in semiconductor structures is simply unknown. In
Ref.~\onlinecite{depalo}, this parameter is uncritically assumed to be
zero, whereas the theoretical results, as shown below, do depend on
the depletion charge density, and even an immeasurably low depletion
density (e.g. $10^9 - 10^{10} cm^{-2}$) changes the results
substantially.

It is well-known that the finite thickness of the 2D electron system
depends on the depletion charge density in the semiconductor
heterostructure. Within the simple variational wavefuntion used in
Ref.~\onlinecite{depalo}, the width parameter `$b$' parameterizing the
quasi-2D thickness is given by $b \equiv (48 \pi m_{\perp} e^2
N^*/\kappa \hbar^2)^{1/3}$, where $m_{\perp}$, $\kappa$ are the
transverse carrier mass and the semiconductor dielectric constant
respectively, and $N^*=N_d + {11 \over 32} N_s$ where $N_d$ is the
depletion charge density and $N_s$ is the 2D carrier density.  Since
$b$ is the key quasi-2D thickness parameter modifying the effective
Coulomb interaction from the strict 2D form of $v(q)=2 \pi e^2/{\kappa
  q}$ to the quasi-2D form of $\tilde{v}(q) = v(q)F(q)$ where $F(q) (<
1)$ is given by Eq.~(2) of Ref.~\onlinecite{depalo}, an accurate
knowledge of $N_d$ is crucial in the determination of the quasi-2D
thickness effect, particularly for low $N_s$ which is the regime of
experimental interest.

\begin{figure}[htbp]
  \centering
  \includegraphics[width=2.5in]{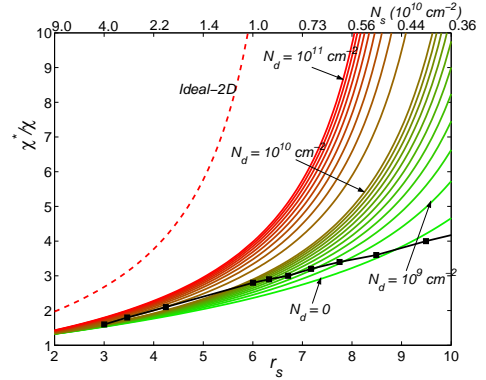}
  \caption{(Color online.) Calculated spin susceptibility
    $\chi^*/\chi$, with $\chi^*$ ($\chi$) the renormalized (Pauli)
    spin susceptibility. The dashed line denotes the ideal-2D result,
    the marked line is the experimental GaAs data from
    Ref.~\onlinecite{zhu}, and the solid lines correspond to $N_d$
    value ranging from $0$ to $10^{10} cm^{-2}$ (in steps of $10^9
    cm^{-2}$) and then to $10^{11} cm^{-2}$ (in steps of $10^{10}
    cm^{-2}$) from bottom to top.  $r_s \equiv (\pi N_s)^{-1/2} /
    a_B$, where $a_B = \kappa \hbar^2 / (m e^2)$, is the usual
    interaction parameter.}
  \label{fig1}
\end{figure}

To emphasize the quantitative importance of the depletion charge
density, we show in Fig.~\ref{fig1} a diagrammatic many-body
calculation of the quasi-2D susceptibility using the so-called
ladder-bubble (i.e. dynamically screened Hartree-Fock or RPA)
approximation~\cite{sus} for a series of values of $N_d = 0 - 10^{11}
cm^{-2}$ for the same 2D GaAs HIGFET structure of
Ref.~\onlinecite{zhu} with which Ref.~\onlinecite{depalo} carried out
their QMC theoretical comparison. The most important feature of
Fig.~\ref{fig1} is that our theory~\cite{sus} also gives ``very good
agreement'' with experimental results if we assume $N_d = 0$ as has
apparently been assumed in Ref.~\onlinecite{depalo}. On the other
hand, any reasonable finite value of $N_d$ leads to increasingly worse
quantitative agreement with experimental results, particularly at
larger values of $r_s$ (i.e. lower values of $N_s$) where the
many-body renormalization is strongest. It fact not only our
theory~\cite{sus}, but also other theoretical
approaches~\cite{dharma}, give ``very good agreement'' with the
experimental data of Ref.~\onlinecite{zhu} if the (unknown) depletion
charge density is assumed to be small (in particular $N_d = 0$) as
done uncritically in Ref.~\onlinecite{depalo}.  We emphasize that the
actual depletion charge density in Ref.~\cite{zhu} is simply not
known, and hence the claim of ``very good agreement''~\cite{depalo}
between QMC results and experiment is meaningless. (Note that in our
work~\cite{sus} we have been very critical about this agreement, and
the ``good agreement'' alone is not the main message of
Ref.~\onlinecite{sus}.)

There are other important physical mechanisms left out of
consideration in Ref.~\onlinecite{depalo} as well: Magneto-orbital
coupling in a finite parallel field~\cite{hwang};
temperature~\cite{galitski}; spin polarization effects~\cite{sus, zhu,
  polarized}; Landau quantization~\cite{smith}. But based on our
discussion of the depletion charge effect alone, it is already safe to
conclude that the ``very good agreement'' between the experimental
data and the QMC theory is accidental. In fact, the agreement between
the QMC theory~\cite{depalo} and experiment is {\it no better} than in
other recent theories~\cite{sus, dharma} of 2D susceptibility.

%%%%%%%%%%%%%%%%%%%%%%%%%%%%%%%%%%%%%%%%%%%%%%%%%%%%%%%%%%%%%%%%%%%%%%%%%%%%%%%

\end{document}